\documentclass[12pt]{article}
\usepackage{graphicx,epsf}

\newcommand\epjc[3]{Eur.\ Phys.\ J.\ C {\bf #1} (#2) #3}

\newcommand\jhep[3]{J.\ High Ener.\ Phys.\ {\bf #1} (#2) #3}
\newcommand\npb[3]{Nucl.\ Phys.\ B {\bf #1} (#2) #3}

\newcommand\plb[3]{Phys.\ Lett.\ B {\bf #1} (#2) #3}
\newcommand\prd[3]{Phys.\ Rev.\ D {\bf #1} (#2) #3}

\newcommand\prl[3]{Phys.\ Rev.\ Lett.\ {\bf #1} (#2) #3}

\newcommand\zpc[3]{Z.\ Phys.\ C {\bf #1} (#2) #3}
\newcommand{\hepph}[1]{{\tt hep-ph/#1}}

\newcommand{\hepex}[1]{{\tt hep-ex/#1}}

\begin{document}

\begin{titlepage}

\begin{flushright}
CLNS~01/1753\\
{\tt hep-ph/0108103}\\[0.2cm]
August 10, 2001
\end{flushright}

\vspace{1.8cm}
\begin{center}
\Large\bf\boldmath
Comments on Color-Suppressed Hadronic $B$ Decays
\unboldmath
\end{center}

\vspace{0.5cm}
\begin{center}
Matthias Neubert and Alexey A. Petrov\\[0.1cm]
{\sl Newman Laboratory of Nuclear Studies, Cornell University\\
Ithaca, NY 14853, USA}
\end{center}

\vspace{1.8cm}
\begin{abstract}
\vspace{0.2cm}\noindent
Recent experimental results on the color-suppressed nonleptonic decays
$\bar B^0\to D^{(*)0}\pi^0$ provide evidence for a failure of the naive 
factorization model and for sizeable relative strong-interaction phases 
between class-1 and class-2 $\bar B\to D^{(*)}\pi$ decay amplitudes. The 
allowed regions for the corresponding ratios of (complex) isospin 
amplitudes and $a_2/a_1$ parameters are determined. The results are 
interpreted in the context of QCD factorization for the related class-1 
amplitudes in the heavy-quark limit.
\end{abstract}

\end{titlepage}

The problem of understanding nonperturbative strong-interaction effects 
in exclusive nonleptonic weak decays of hadrons has always been a 
challenge to theorists. Only in a few cases model-independent results 
based on controlled expansions in QCD can be obtained. In the absence of 
a quantitative theoretical description various attempts have been made 
to obtain simple, predictive parameterizations of decay amplitudes based 
on simple phenomenological assumptions. 

The most common of these approaches is the ``naive'' (or ``generalized'')
factorization model, in which the decay amplitudes are estimated by 
replacing hadronic matrix elements of four-quark operators in the 
effective weak Hamiltonian by products of current matrix elements 
determined in terms of meson decay constants and semileptonic form 
factors. ``Nonfactorizable'' strong-interaction effects are 
parameterized by phenomenological coefficients $a_i$, which depend on the 
color and Dirac structure of the operators but otherwise are postulated 
to be universal \cite{BSW,NRSX,Dean93}. One distinguishes class-1 and 
class-2 decay topologies, which refer to the cases where a charged 
(class-1) or a neutral (class-2) final-state meson can be produced from 
the quarks contained in the four-quark operators of the effective 
Hamiltonian. For instance, in decays based on the quark transition 
$b\to c\bar u d$ mesons with quark content $(\bar u d)$ or $(\bar u c)$ 
can be produced in that way. The decay $\bar B^0\to D^+\pi^-$ is a 
class-1 process, in which the pion can be generated at the weak vertex, 
whereas $\bar B^0\to D^0\pi^0$ is a class-2 process, in which the $D^0$ 
meson can be directly produced. The corresponding amplitudes are then 
expressed as 
\begin{eqnarray}\label{ampl}
   {\cal A}(\bar B^0\to D^+\pi^-) &=& i\,\frac{G_F}{\sqrt 2}\,
    V_{cb} V_{ud}^*\,(m_B^2-m_D^2)\,f_\pi\,F_0^{B\to D}(m_\pi^2)\,
    a_1(D\pi) \,, \nonumber\\
   \sqrt 2\,{\cal A}(\bar B^0\to D^0\pi^0) &=& i\,\frac{G_F}{\sqrt 2}\,
    V_{cb} V_{ud}^*\,(m_B^2-m_\pi^2)\,f_D\,F_0^{B\to\pi}(m_D^2)\,
    a_2(D\pi) \,, 
\end{eqnarray}
where $F_0^{B\to M}(q^2)$ are $B\to M$ form factors at momentum transfer 
$q^2$. In other processes such as $B^-\to D^0\pi^-$ both topologies can 
contribute and interfere. (Such processes are sometimes called class-3 
decays.) In fact, isospin symmetry implies that
\begin{equation}
   {\cal A}(B^-\to D^0\pi^-) = {\cal A}(\bar B^0\to D^+\pi^-)
   + \sqrt 2\,{\cal A}(\bar B^0\to D^0\pi^0) \,.
\end{equation}
The large-$N_c$ counting rules of QCD show that $a_1(D\pi)=O(1)$ and 
$a_2(D\pi)=O(1/N_c)$, which is why the class-2 decays are often referred 
to as ``color suppressed''. 

In the naive factorization model one postulates that for a large class 
of energetic, two-body (or quasi two-body) $B$ decays the coefficients 
$a_1$ and $a_2$ are process-independent phenomenological parameters. 
These parameters are assumed to be real, ignoring the possibility of 
relative strong-interaction phases between class-1 and class-2 
amplitudes. Surprisingly, despite their crudeness these assumptions 
seemed to be supported by experimental data \cite{NRSX,Dean93}. Within 
errors, the class-1 decays $\bar B^0\to D^{(*)+} M^-$ with 
$M=\pi,\rho,a_1,D_s,D_s^*$ can be described using a universal value 
$|a_1|\approx 1.1\pm 0.1$, whereas the class-2 decays 
$\bar B\to\bar K^{(*)} M$ with $M=J/\psi,\psi(2S)$ suggest a nearly 
universal value $|a_2|\approx 0.2$--0.3 (which is more uncertain due to 
the uncertainty in the $B\to K^{(*)}$ form factors). Moreover, the 
class-3 decays $B^-\to D^{(*)0} M^-$ with $M=\pi,\rho$, which are 
sensitive to the interference of the two decay topologies, could be 
explained by a real, positive ratio $a_2/a_1\approx 0.2$--0.3, which 
seemed to agree with the determinations of $|a_1|$ and $|a_2|$ from 
other decays. The missing link in this line of argument was a direct 
measurement of $|a_2|$ in the related class-2 decays 
$\bar B^0\to D^{(*)0} M^0$.

Recently, the idea of factorization in the class-1 decays 
$\bar B^0\to D^{(*)+} L^-$, where $L$ is a {\em light\/} meson, was put 
on a more rigorous footing. It was shown that the corresponding decay 
amplitudes can be systematically calculated in QCD in the limit where 
the decaying $b$ quark is considered a heavy quark 
\cite{BBNS1,BBNS2,BPS01} (see \cite{PoWi,DuGr,DoPe} for related earlier 
work). To leading order in $\Lambda/m_b$ (with $\Lambda$ a typical 
hadronic scale), but to all orders of perturbation theory, 
nonfactorizable strong-interaction effects can be described in terms of 
convolutions of hard-scattering kernels with the leading-twist 
light-cone distribution amplitude of the light meson $L$. The resulting 
QCD factorization formula allows us to compute the magnitude and phase 
of the parameters $a_1(D^{(*)}L)$ systematically up to power corrections 
in $\Lambda/m_b$. The values of $a_1$ in different class-1 decays are 
not universal, but the process-dependent corrections turn out to be 
numerically small \cite{BBNS2}. Several types of power corrections to 
the $a_1$ parameters have been estimated and found to be small 
\cite{BBNS2,Khod,Burrell,us}. Hence, for the cases where $L$ is a light 
meson there is now a solid theoretical understanding of the 
near-universal value $|a_1|\approx 1.1$ observed experimentally.

On the other hand, if the charm quark is treated as a heavy quark, then 
the QCD factorization formula does {\em not\/} apply for the class-2 
decays $\bar B^0\to D^{(*)0} L^0$, and so the magnitude and phase of the 
$a_2(D^{(*)}L)$ parameters are not calculable. The only nontrivial 
prediction in this case is that the class-2 amplitudes are power 
suppressed with respect to the corresponding class-1 amplitudes 
\cite{BBNS2}. The apparent universality of the $|a_2|$ values extracted 
from experiment, and the absence of sizeable relative strong-interaction 
phases between the various $a_1$ and $a_2$ parameters suggested by the
data, remained a theoretical puzzle. (Even before the advent of QCD 
factorization various authors had presented arguments against the 
universality of nonfactorizable effects in class-2 decays; see, e.g.,  
\cite{Cheng,Soares,Pene}.) New experimental data announced by the CLEO 
and Belle Collaborations \cite{CLEO,Belle} change the picture 
significantly, in a way that is entirely in line with QCD expectations. 
As we will illustrate below, these data provide compelling evidence for 
{\em process-dependent\/} $a_2$ values, and for large {\em relative 
strong-interaction phases\/} between related class-1 and class-2 
amplitudes. This shows that the ``naive'' factorization model is too 
simple to account for the data. 

Because QCD factorization cannot be justified for the class-2 decays, it 
is in some sense misleading to parameterize the $\bar B^0\to D^0\pi^0$ 
decay amplitude as done in (\ref{ampl}). Although the naive factorization
contribution $a_2=C_2+\zeta\,C_1$ (with $\zeta$ a nonperturbative 
parameter of order $1/N_c$) is certainly present, it is not a leading 
contribution to the decay amplitude in any consistent limit of QCD. For 
instance, there exists a weak annihilation contribution to $a_2$ which 
scales like a power of $m_b/\Lambda$ in the heavy-quark limit and thus 
formally dominates over the naive factorization piece. (Weak annihilation 
is mentioned here only as an example of a leading contribution to the 
class-2 amplitude. Model calculations suggest that the annihilation 
contribution in $\bar B\to D\,\pi$ decays is nevertheless small 
\cite{BBNS2}. Another example of a leading contribution is 
charge-exchanging rescattering from the dominant class-1 channel 
\cite{resc,BlHa}.) Also, for class-2 decays naive factorization does not 
emerge in the large-$N_c$ limit. It is then more appropriate to employ 
an alternative parameterization of the decay amplitudes in terms of 
isospin amplitudes $A_{1/2}$ and $A_{3/2}$ corresponding to transitions 
into $D\,\pi$ final states with $I=\frac12$ and $\frac32$, respectively. 
It is given by
\begin{eqnarray}
   {\cal A}(\bar B^0\to D^+\pi^-) &=& \sqrt{\frac13}\,A_{3/2}
    + \sqrt{\frac23}\,A_{1/2} 
    \,, \nonumber\\
   \sqrt 2\,{\cal A}(\bar B^0\to D^0\pi^0) &=& \sqrt{\frac43}\,A_{3/2}
    - \sqrt{\frac23}\,A_{1/2} 
    \,, \nonumber\\[0.2cm]
   {\cal A}(B^-\to D^0\pi^-) &=& \sqrt3\,A_{3/2} 
    \,.
\end{eqnarray}
An identical decomposition holds for other decays such as 
$\bar B\to D^*\pi$ and $\bar B\to D^{(*)}\rho$. It follows from QCD 
factorization that the ratio of isospin amplitudes is \cite{BBNS2}
\begin{equation}\label{Aratio}
   \frac{A_{1/2}}{\sqrt 2\,A_{3/2}} = 1 + O(\Lambda/m_Q) \,,
\end{equation}
which also implies that the relative strong-interaction phase
$\delta_{1/2}-\delta_{3/2}=O(\Lambda/m_Q)$. Here $m_Q$ represents either
one of $m_c$ and $m_b$. Note that the corrections to the ``1'' in 
(\ref{Aratio}) are also formally suppressed by a power of $1/N_c$, and 
it has been argued that perhaps this color suppression may be more 
relevant to factorization than the heavy-quark limit \cite{Zoltan}. 
However, an identical $1/N_c$ argument would apply to other nonleptonic 
decays such as $D\to\bar K^{(*)}\pi$ and $K\to\pi\pi$, for which color 
suppression is clearly not operative. Apart from trivial substitutions 
of quark flavors, the only difference between, say, 
$\bar B\to D^{(*)}\pi$ and $D\to\bar K^{(*)}\pi$ decays is the larger 
energy release in the decay of a heavy $b$ quark, which leads to color 
transparency and thus is the basis of QCD factorization.

\begin{table}[t]
\centerline{\parbox{14cm}{\caption{\label{tab:scal} 
Heavy-quark scaling laws in the schemes where the charm quark is treated 
as a heavy quark (left) or as a light quark (right)}}}
\vspace{0.2cm}
\begin{center}
{\tabcolsep=0.5cm
\begin{tabular}{|c|cc|}
\hline\hline
 & $m_c\sim m_b\sim m_Q$ & $m_c\sim\Lambda\ll m_b$ \\
\hline
$|A_{1/2}/\sqrt2 A_{3/2}|$ & $1+O(\Lambda/m_Q)$ & $O(1)$ \\
$\delta_{1/2}-\delta_{3/2}$ & $O(\Lambda/m_Q)$ & $O[\alpha_s(m_b)]$ \\
\hline\hline
\end{tabular}}
\end{center}
\end{table}

The deviation of the ratio $A_{1/2}/(\sqrt 2\,A_{3/2})$ from 1 is a 
measure of the departure from the heavy-quark limit. When contemplating 
about the expected magnitude of this effect, it is important to realize 
that the {\em power suppression\/} of the corrections in (\ref{Aratio}) 
relies on the heaviness of the charm quark, not only the $b$ quark. In 
order to illustrate this fact it is instructive to consider two 
different power-counting schemes for the heavy-quark expansion. The most 
natural scheme, which underlies our discussion so far, is to consider 
both beauty and charm as heavy quarks with their mass ratio $m_c/m_b$ 
fixed in the heavy-quark limit. Then the deviation of 
$A_{1/2}/(\sqrt 2\,A_{3/2})$ from 1 is power suppressed in $\Lambda/m_Q$, 
where $m_Q\sim m_b\sim m_c$. It is not calculable without model 
dependence. Alternatively, one may consider the charm quark as a light 
quark with its mass kept fixed in the heavy-quark limit \cite{BBNS2}. 
In such a scheme the class-2 amplitude becomes calculable and of leading 
power in the limit $m_b\gg\Lambda$ \cite{BBNS3}. The leading deviation 
of the isospin amplitude ratio from 1 is then computable in terms of 
Wilson coefficient functions $C_i(m_b)$ and short-distance corrections 
proportional to $\alpha_s(m_b)$. The scaling of the relevant quantities 
in these two versions of the heavy-quark limit is summarized in 
Table~\ref{tab:scal}. The conventional scheme is, perhaps, somewhat 
closer to reality. However, considering that the charm quark is not
very heavy in the real world, we may expect a sizeable deviation of
the amplitude ratio from 1. 

\begin{table}[t]
\centerline{\parbox{14cm}{\caption{\label{tab:results} 
Experimental data for the $\bar B\to D^{(*)}\pi$ and 
$D\to\bar K^{(*)}\pi$ branching ratios (in units of $10^{-3}$), isospin 
amplitudes, and related quantities}}}
\vspace{0.2cm}
\begin{center}
{\tabcolsep=0.16cm
\begin{tabular}{|c|cc||c|cc|}
\hline\hline
 & $\bar B\to D\,\pi$ & $\bar B\to D^*\pi$ &
 & $D\to\bar K\pi$ & $D\to\bar K^*\pi$ \\
\hline
$\bar B^0\to D^{(*)+}\pi^-$ & $3.0\pm 0.4$ & $2.76\pm 0.21$ &
$D^0\to K^{(*)-}\pi^+$ & $38.3\pm 0.9$ & $50\pm 4$ \\
$\bar B^0\to D^{(*)0}\pi^0$ & $0.27\pm 0.05$ & $0.17\pm 0.05$ &
$D^0\to\bar K^{(*)0}\pi^0$ & $21.1\pm 2.1$ & $31\pm 4$ \\
$B^-\to D^{(*)0}\pi^-$ & $5.3\pm 0.5$ & $4.6\pm 0.4$ &
$D^+\to\bar K^{(*)0}\pi^+$ & $28.9\pm 2.6$ & $19.0\pm 1.9$ \\
\hline
$|A_{1/2}/\sqrt2\,A_{3/2}|$ & $0.70\pm 0.11$ & $0.72\pm 0.08$ &
$|A_{1/2}/\sqrt2\,A_{3/2}|$ & $2.71\pm 0.14$ & $3.97\pm 0.25$ \\
$|\delta_{1/2}-\delta_{3/2}|$ & $(27\pm 7)^\circ$ & $(21\pm 8)^\circ$ &
$|\delta_{1/2}-\delta_{3/2}|$ & $(90\pm 6)^\circ$
 & $(104\pm 13)^\circ$ \\
\hline
$x\,|a_2/a_1|$ & $0.42\pm 0.05$ & $0.35\pm 0.05$ & 
$x\,|a_2/a_1|$ & $1.05\pm 0.05$ & $1.11\pm 0.08$ \\ 
$\mbox{arg}(a_2/a_1)$ & $(56\pm 20)^\circ$ & $(51\pm 20)^\circ$ &
$\mbox{arg}(a_2/a_1)$ & $(149\pm 2)^\circ$ & $(160\pm 2)^\circ$ \\
$x\,a_2^{\rm eff}/a_1^{\rm eff}$ & $0.25\pm 0.12$ & $0.23\pm 0.08$ &
$x\,a_2^{\rm eff}/a_1^{\rm eff}$ & $-0.53\pm 0.02$
 & $\!\!-0.66\pm 0.02$ \\
$a_2^{\rm eff}/a_1^{\rm eff}$ & $\approx 0.28$ & $\approx 0.25$ &
$a_2^{\rm eff}/a_1^{\rm eff}$ & $\approx -0.44$ & $\approx -0.35$ \\
\hline\hline
\end{tabular}}
\end{center}
\end{table}

We now investigate to what extent the prediction (\ref{Aratio}) is 
supported by the data on $\bar B\to D^{(*)}\pi$ decays. For comparison,
it will be instructive to analyze the related charm decays 
$D\to\bar K^{(*)}\pi$ in parallel. The upper portion in 
Table~\ref{tab:results} summarizes the experimental data on the various 
branching ratios. The color-suppressed $B$ decays have just been observed
for the first time experimentally \cite{CLEO,Belle}. The preliminary 
results for the $\bar B^0\to D^0\pi^0$ branching ratio are 
$(2.6\pm 0.3\pm 0.6)\times 10^{-4}$ (CLEO) and 
$(2.9_{\,-0.3}^{\,+0.4}\pm 0.6)\times 10^{-4}$ (Belle), while those
for the $\bar B^0\to D^{*0}\pi^0$ branching ratio are 
$(2.0\pm 0.5\pm 0.7)\times 10^{-4}$ (CLEO) and 
$(1.5_{\,-0.5\,-0.4}^{\,+0.6\,+0.3})\times 10^{-4}$ (Belle). We have 
averaged these results to obtain the entries shown in the table. All 
other numbers are taken from \cite{PDG}. By combining the measurements 
of the three branching ratios for each mode, taking into account the 
lifetime ratios $\tau(B^-)/\tau(\bar B^0)=1.068\pm 0.016$ \cite{Blife} 
and $\tau(D^+)/\tau(D^0)=2.547\pm 0.036$ \cite{PDG}, one can extract the 
magnitude and phase of the ratio of isospin amplitudes. The results are 
shown in the middle portion of the table. A clear qualitative difference 
between beauty and charm decays emerges. Whereas the relative phases are 
close to maximal in $D$ decays and the amplitude ratios are far from the 
asymptotic value in (\ref{Aratio}), the corrections to the heavy-quark 
limit appear to be much smaller in $B$ decays. 

\begin{figure}[t]
\begin{center}
\includegraphics[width=0.95\textwidth]{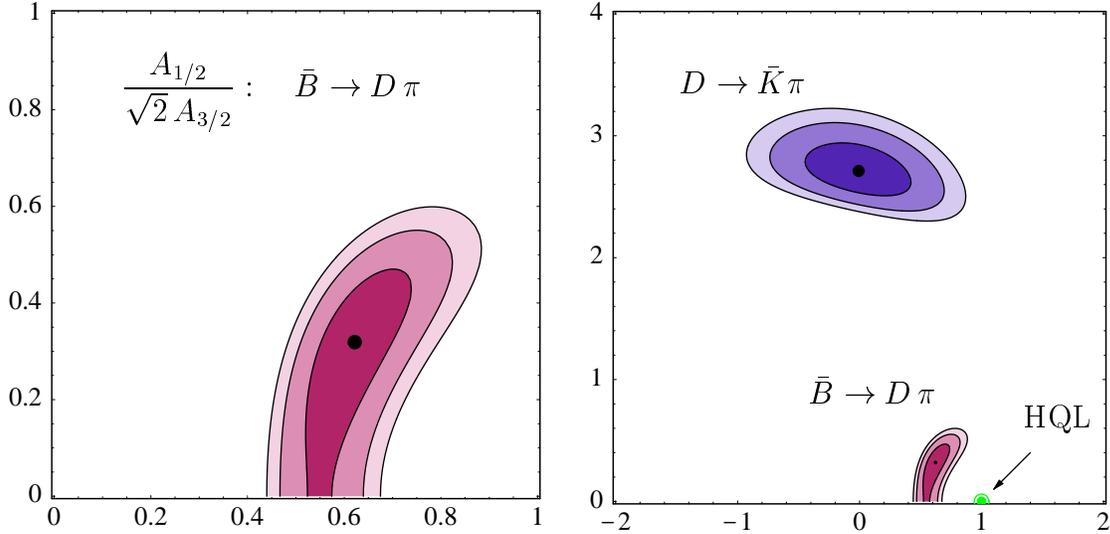}
\vspace{0.2cm}\\
\parbox{14cm}{\caption{\label{fig:1}
Allowed regions in the complex $A_{1/2}/(\sqrt 2\,A_{3/2})$ plane 
obtained at 68\%, 95\% and 99\% confidence level. The sign of the 
imaginary part is undetermined by the data. The black dots show the 
central values.}}
\end{center}
\end{figure}

A more careful analysis of the experimental data for $\bar B\to D\,\pi$ 
and $D\to\bar K\pi$ decays is shown in Figure~\ref{fig:1}, which gives 
the allowed regions for the ratio $A_{1/2}/(\sqrt 2\,A_{3/2})$ obtained 
at different confidence levels. To derive these regions we find, for 
each value of the isospin amplitude ratio, the minimum of the $\chi^2$ 
function for the three measured branching ratios. We then plot contours 
of minimum $\chi^2$ in the complex plane corresponding to a given 
confidence level (for two degrees of freedom). Very similar constraints 
can be derived for $\bar B\to D^*\pi$ and $D\to\bar K^*\pi$ decays. 
Note that even at the level of one standard deviation (68\% confidence 
level) the relative strong phase of the isospin amplitudes for $B$ 
decays shown in the left plot may be zero, in contrast with the naive 
error propagation in Table~\ref{tab:results}. In the right plot we 
compare the results in $D$ and $B$ decays and also indicate the value 
corresponding to the strict heavy-quark limit (HQL). We conclude that 
the data on the ratio of isospin amplitudes in $\bar B\to D^{(*)}\pi$ 
decays is compatible with the heavy-quark scaling laws discussed earlier.
The comparison of charm to beauty decays shows a clear progression 
towards the heavy-quark limit as the mass of the decaying quark 
increases. The remaining deviation of the amplitude ratio from 1 is 
compatible with a correction whose suppression is governed by the large 
charm-quark mass.

\begin{figure}[t]
\begin{center}
\includegraphics[width=0.95\textwidth]{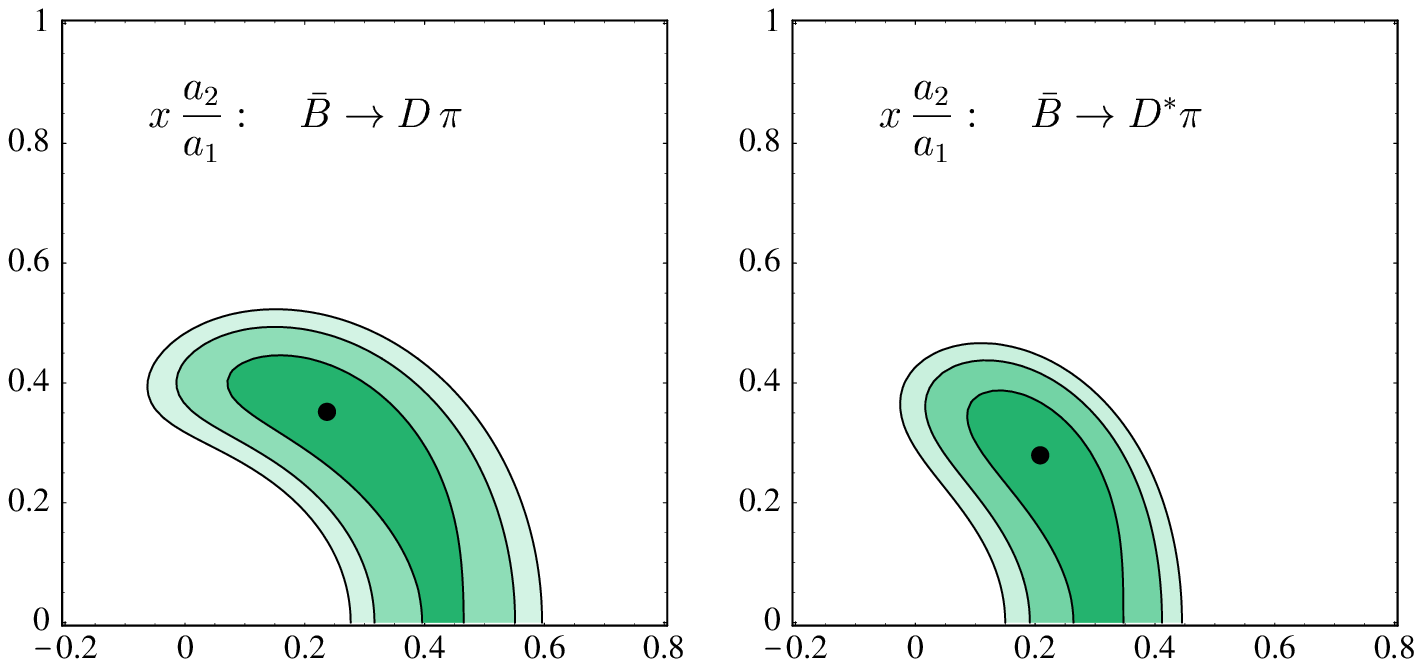}
\vspace{0.2cm}\\
\parbox{14cm}{\caption{\label{fig:2}
Allowed regions in the complex $x\,a_2/a_1$ plane for 
$\bar B\to D^{(*)}\pi$ decays, obtained at 68\%, 95\% and 99\% 
confidence level. The sign of the imaginary part is undetermined by the
data. The black dots show the central values.}}
\end{center}
\end{figure}

We have mentioned earlier that the parameterization of the class-2 
amplitude in (\ref{ampl}) is somewhat misleading, because the naive 
factorization contribution is in no way the leading term in a controlled
expansion of the decay amplitude. Nevertheless, it is instructive to 
extract the parameter $a_2$ defined via the second relation in 
(\ref{ampl}) and compare it with the values of the $a_2$ parameters 
obtained from other decays, such as $\bar B\to\bar K^{(*)} J/\psi$. The 
ratios of the $\bar B\to D^{(*)}\pi$ branching ratios can be expressed 
in terms of the ratios $x\,a_2/a_1$, where
\begin{eqnarray}\label{xrat}
   x(D\pi) &=& \frac{(m_B^2-m_\pi^2)\,f_D\,F_0^{B\to\pi}(m_D^2)}
                    {(m_B^2-m_D^2)\,f_\pi\,F_0^{B\to D}(m_\pi^2)}
    \approx 0.9 \,, \nonumber\\
   x(D^*\pi) &=& \frac{f_{D^*} F_+^{B\to\pi}(m_{D^*}^2)}
                      {f_\pi\,A_0^{B\to D^*}(m_\pi^2)}
    \approx 0.9 \,.
\end{eqnarray}
The numerical values have been obtained using $f_D\approx 200$\,MeV, 
$f_{D^*}\approx 230$\,MeV, and 
$F_0^{B\to\pi}(m_D^2)/F_0^{B\to D}(m_\pi^2)\approx 
F_+^{B\to\pi}(m_{D^*}^2)/A_0^{B\to D^*}(m_\pi^2)\approx 0.5$. (The 
corresponding quantities for $D$ decays may be estimated using the BSW 
model \cite{BSW}, with the result that $x(\bar K\pi)\approx 1.2$ and
$x(\bar K^*\pi)\approx 1.9$.) Figure~\ref{fig:2} shows the corresponding 
allowed regions for $x\,a_2/a_1$ obtained at different confidence levels. 
The central values are also shown in Table~\ref{tab:results}. The data 
prefer values $|a_2|\approx 0.4$--0.5, which are larger by almost a 
factor 2 than those obtained from $\bar B\to\bar K^{(*)} J/\psi$ decays 
(see above).\footnote{Preliminary Belle data \protect\cite{Belle} on 
related color-suppressed decays support this conclusion. Using the 
form-factor model of \protect\cite{NRSX}, we find that the decays 
$\bar B^0\to D^{(*)0}\eta$ and $\bar B^0\to D^{(*)0}\omega$ have $|a_2|$ 
values between 0.4 and 0.5, with experimental errors of about 0.1 and a 
theoretical uncertainty of about 30\%.}
Moreover, the best fits prefer phases of about $50^\circ$ in the two 
cases (with large errors), suggesting that strong final-state 
interactions cannot be neglected and lead to a nontrivial relative phase 
between class-1 and class-2 amplitudes. Both observation are in conflict 
with the assumptions underlying the naive factorization model. It now 
appears that the $a_2$ coefficients of the class-2 amplitudes are 
nonuniversal, with magnitudes and phases that may be rather different 
for different types of decays. More precise data will be required to 
fully explore the pattern of QCD effects in the class-2 and class-3 
amplitudes.

We should mention that also in $D\to\bar K^{(*)}\pi$ decays the extracted 
values of $a_2/a_1$ have large phases and are larger in magnitude than 
those extracted from other $D$ decays (see Table~\ref{tab:results}). This 
``failure'' of naive factorization is usually attributed to the strong 
final-state interactions caused by nearby resonances (as signaled by the 
fact that $|\delta_{1/2}-\delta_{3/2}|\approx 90^\circ$ in these decays). 
It is then argued that one can only expect to correctly predict the 
{\em magnitudes\/} of the isospin amplitudes but not their relative phase
\cite{NRSX}. The ratio of these magnitudes is determined by the ratio of
the real, effective $a_1^{\rm eff}$ and $a_2^{\rm eff}$ parameters of the
naive factorization model via the relation
\begin{equation}
   \frac{|A_{1/2}|}{\sqrt2\,|A_{3/2}|}
   = \frac{2-x\,a_2^{\rm eff}/a_1^{\rm eff}}
          {2(1+x\,a_2^{\rm eff}/a_1^{\rm eff})} \,.
\end{equation}
The effective parameters so determined are shown in the lower portion of 
Table~\ref{tab:results}. In the case of $D$ decays, the physical picture 
underlying this approach is that of predominantly elastic final-state 
interactions, which mix the various $\bar K^{(*)}\pi$ final states and 
thereby changes the phases but not the magnitudes of the isospin 
amplitudes. While this assumption may be questioned even in the case of 
charm decays \cite{Adam}, is it clearly not justifiable for decays of 
$B$ mesons, in which rescattering is predominantly inelastic 
\cite{BBNS2,resc}. Therefore, we believe it is a coincidence that the 
``effective'' $a_2^{\rm eff}/a_1^{\rm eff}$ ratios are close to the 
expectations of the naive factorization model.

Finally, we like to comment on the observation that the ratios $x$ in 
(\ref{xrat}), which govern the relative strength of the class-2 and 
class-1 amplitudes in naive factorization, exhibit large violations of 
the scaling $x\sim(\Lambda/m_Q)^2$ expected in the heavy-quark limit. 
Does this imply a failure of QCD factorization in hadronic $B$ decays? 
We believe the answer to this question is negative. Consider first the 
conventional case where the charm quark is treated as a heavy quark. 
Then the ratios $x$ arise only in naive factorization. The fact that 
they are not numerically suppressed reflects the well-known failure of 
the conventional heavy-quark expansion for heavy-light form factors and 
decay constants, i.e., the empirical fact that the ratios 
\begin{equation}\label{viol}
   \frac{f_D}{f_\pi}\approx 1.5 \quad [\sim (\Lambda/m_Q)^{1/2}] \,,
    \qquad
   \frac{F_0^{B\to\pi}(m_D^2)}{F_0^{B\to D}(m_\pi^2)}\approx 0.5
    \quad [\sim (\Lambda/m_Q)^{3/2}] 
\end{equation}
do not scale as expected from the heavy-quark limit of QCD. (The reason 
for this failure may be related to the ``smallness'' of $f_\pi$, which 
in turn reflects the smallness of the light quark masses via the 
relation $m_\pi^2\,f_\pi^2=-2(m_u+m_d)\,\langle\bar q q\rangle$.) 
However, as 
we have argued earlier the factorized class-2 contributions appearing 
in the numerator of the ratios $x$ are not a leading contribution to the 
class-2 amplitudes in the heavy-quark limit. Other contributions exist 
that are parametrically larger. It is therefore not clear to what extent 
the large scaling violations in (\ref{viol}) are relevant to the class-2 
amplitudes. In the opposite limit where the charm quark is considered a 
light quark the naive factorization contribution is the leading 
contribution to the class-2 amplitudes. In this limit also the ratios 
$x$ are of 
leading order in the heavy-quark expansion, which is consistent with the 
numerical values $x(D^{(*)}\pi)\approx 0.9$. So there is no evidence for 
a failure of the heavy-quark expansion either.

In summary, we have argued that new experimental results on the 
color-suppressed nonleptonic decays $\bar B^0\to D^{(*)0}\pi^0$ provide 
evidence for a failure of the naive factorization model and for sizeable 
relative strong-interaction phases between class-1 and class-2 
$\bar B\to D^{(*)}\pi$ decay amplitudes. This resolves a long-standing 
puzzle created by the apparent universality and small rescattering 
phases of the class-2 parameters $a_2$ in $B$ decays. The new data 
suggest that the $a_2$ parameters in different types of decays such as
$\bar B\to D^{(*)}\,\pi$ and $\bar B\to\bar K^{(*)} J/\psi$ differ by
almost a factor 2 in magnitude, indicating a strong nonuniversality of
nonfactorizable effects. This is in agreement with theoretical 
expectations based on the heavy-quark expansion. We find that the 
size of corrections to the heavy-quark limit seen in the data is 
compatible with the expectation that the suppression of the corrections 
is governed by the large charm-quark mass. We urge our experimental 
colleagues to produce more precise data on a large variety of hadronic 
$B$ decays. This will help to explore in detail the pattern of QCD 
effects in the class-2 and class-3 amplitudes, as well as to further 
establish the validity of QCD factorization (and hence the applicability 
of the heavy-quark limit) for class-1 decays.

\vspace{0.3cm}
\noindent
{\it Acknowledgement:\/}
This work was supported in part by the National Science Foundation.

\vspace{0.3cm}
\noindent
{\it Note added:\/}
While this Letter was in writing the paper \hepph{0107257} by Z.-z.\ Xing
appeared, in which a similar analysis is carried out but different 
conclusions are obtained.

\end{document}